\documentclass[a4paper]{cas-dc}
\usepackage[numbers]{natbib}
\usepackage{cuted}\stripsep-2pt
\usepackage{subcaption,mathtools}

\def\tsc#1{\csdef{#1}{\textsc{\lowercase{#1}}\xspace}}
\tsc{WGM}
\tsc{QE}
\tsc{EP}
\tsc{PMS}
\tsc{BEC}
\tsc{DE}

\begin{document}
\let\WriteBookmarks\relax
\def\floatpagepagefraction{1}
\def\textpagefraction{.001}
\shorttitle{Multi-Gaussian Model for Intrabeam Scattering}
\shortauthors{W.Wu et~al.}

\title [mode = title]{Multi-Gaussian Model for Intrabeam Scattering}

\author[1]{Wenxuan Wu}

\affiliation[1]{organization={Department of Engineering Physics, Tsinghua University},
                city={Beijing},
                postcode={100084}, 
                country={China}}

\author[1,3]{Zhilong Pan}

\author[1]{Chuanxiang Tang}

\affiliation[2]{organization={Institute of Advanced Study, Tsinghua University},
                postcode={100084}, 
                city={Beijing},
                country={China}}

\affiliation[3]{organization={National Engineering Research Center of Dangerous Articles and Explosives Detection Technologies},
                postcode={100084}, 
                city={Beijing},
                country={China}}

\author[2]{Xiujie Deng}

\begin{abstract}
Conventional analytical models of intrabeam scattering (IBS), such as the Piwinski model and Bjorken–Mtingwa model, generally rely on the assumption of a Gaussian beam distribution, limiting their applicability to non-Gaussian cases. In this work, we develop a multi-Gaussian model to extend analytical IBS calculations to a broader class of non-Gaussian phase-space distributions. The proposed approach represents a non-Gaussian distribution as a superposition of Gaussian components sharing a common covariance matrix but having different mean values. We derived a generalized model for evaluating the momentum second-moment evolution rate and the IBS growth rate. Furthermore, for equally spaced Gaussian components, we developed an infinite bunch train model and introduce a practical truncation rule to reduce computational costs. The model is applied to a representative steady-state microbunching (SSMB) storage ring, where longitudinal overlap between neighboring microbunches is shown to enhance IBS growth rates in all three directions. The proposed framework provides a practical approach to evaluating IBS in non-Gaussian beam distributions and offers a basis for further development toward a more general non-Gaussian IBS theory.
\end{abstract}


\maketitle

\section{\label{sec:1}Introduction}
Intrabeam scattering (IBS) is a fundamental collective effect in accelerator physics. It originates from Coulomb scattering among charged particles within a beam and leads to the growth of beam emittances in storage rings. In modern accelerators, achieving high beam brightness requires both low emittances and high beam currents. Under these conditions, IBS can become a significant limitation to beam performance and must be carefully evaluated.

In our recent studies of steady-state microbunching (SSMB) storage rings, IBS has been identified as an important factor affecting the achievable performance.\cite{deng2023ssmb} The fundamental concept of SSMB was first proposed by Ratner and Chao \cite{ratner2010steady}. By storing microbunched electron beams in a storage ring, SSMB aims to combine the advantages of free-electron lasers (FELs), which provide high-intensity and highly coherent radiation, with those of synchrotron radiation (SR) sources, which offer high repetition rates. Achieving this goal requires ultrashort bunch lengths, particularly when radiation at short wavelengths is desired. Several SSMB operating regimes have been proposed in recent years.\cite{Deng2024SteadystateMB} Regardless of the specific regime, accurate evaluation of IBS is essential for assessing the feasibility and performance of SSMB storage rings.

Both analytical models and numerical methods are commonly employed to calculate IBS effect. Conventional analytical treatments generally assume a Gaussian beam distribution. Based on the classical theory of Coulomb scattering, Piwinski first developed an analytical formulation for IBS \cite{piwinski2005intra}. Subsequently, Bjorken and Mtingwa introduced the B–M model \cite{bjorken1982intrabeam}, which provides a more comprehensive treatment of IBS in storage rings. These formulations have become standard tools for evaluating IBS-induced emittance growth. However, their analytical evaluation is fundamentally tied to the single-Gaussian assumption, which requires the phase space distribution to be Gaussian distribution. This assumption can limit their accuracy when the actual beam distribution deviates significantly from Gaussian.

Non-Gaussian beam distributions are encountered in various accelerator applications, including free-electron lasers and heavy-ion accelerators. Several studies have therefore investigated analytical approaches to non-Gaussian IBS. A straightforward strategy is to divide a non-Gaussian beam into multiple longitudinal slices and treat the distribution within each slice as Gaussian. This approach has been implemented in accelerator simulation codes such as ELEGANT \cite{xiao2015intra}. However, its applicability relies on the assumption that the phase space distribution within each longitudinal slice is adequately described by a Gaussian distribution. Other analytical approaches\cite{meshkov2006physics,papadopoulou2020impact} have also been developed, but their applicability remains largely restricted to distributions that do not deviate substantially from Gaussian. Consequently, analytical treatments of IBS for general non-Gaussian distributions remain limited. Numerical approaches based on the Fokker-Planck equation \cite{zenkevich2006new} and Monte Carlo techniques \cite{alekseev1999numerical}, on the other hand, offer greater flexibility in handling non-Gaussian distributions, but often require substantial computational costs, particularly in large-scale or repetitive simulations.

Motivated by these limitations, we propose a multi-Gaussian analytical model for evaluating IBS in some non-Gaussian cases. The key observation is that many non-Gaussian distributions encountered in storage rings, such as flat-top profiles \cite{byrd2001harmonic,xiao2009intrabeam}, can be approximated by a superposition of a finite number of Gaussian components sharing the same covariance matrix but having different mean values. Building upon the established single-Gaussian IBS model, the proposed approach extends analytical IBS calculations to a broader class of non-Gaussian distributions while retaining the computational efficiency and physical interpretability of the conventional formulation.

To demonstrate the applicability of the multi-Gaussian model, we consider a particular SSMB operating regime, namely the longitudinal weak-focusing regime. This regime provides a useful example because its beam dynamics are relatively straightforward to understand, while exhibiting features that distinguish it from conventional storage rings. Previous studies \cite{Pans_work} have discussed the necessity of employing a fully coupled six-dimensional (6D) IBS model \cite{kubo2001intrabeam,nash2006analytical} rather than the original B-M formulation for this type of storage ring. Nevertheless, even the fully coupled 6D model does not fully capture the effects associated with overlapping microbunches. In certain regions of the storage ring, the local bunch length can exceed the longitudinal separation between adjacent bunch centers due to transverse-longitudinal coupling. Consequently, neighboring bunches overlap in the longitudinal direction, and scattering events can occur between particles belonging to different microbunches. These scattering contributions can enhance the overall IBS rate and are not accounted for by the conventional single-Gaussian treatment. The details of this effect and its quantitative evaluation will be discussed in Section IV.

The remainder of this paper is organized as follows. Section II briefly reviews the single-Gaussian IBS model. Section III introduces the multi-Gaussian model and presents its mathematical formulation. Section IV provides a brief introduction to SSMB and applies the proposed model to a representative SSMB storage ring, followed by a comparison with the fully coupled 6D B-M model. Finally, Section V summarizes the main conclusions and discusses possible extensions toward a more general theoretical framework for non-Gaussian IBS.

\section{Single-Gaussian model}
To establish the theoretical foundation for the multi-Gaussian IBS model, we first review the conventional single-Gaussian formulation, which is based on the modified Bjorken–Mtingwa (B–M) model (fully coupled 6D IBS model). This section provides a concise summary of the relevant theoretical framework and introduces the key quantities that will be generalized in the following section.

The IBS growth rate in the single-Gaussian model is defined in the same way as in the Piwinski formulation:
\begin{equation}
    \frac{1}{\tau_a}=\frac{1}{2\langle g_a\rangle}\frac{d\langle g_a\rangle}{dt}\label{time}
\end{equation}
where $g_a$ represents one of the three invariants of the storage ring ($a=x,y,z$). In a fully decoupled storage ring, these invariants correspond to the conventional emittances ($\epsilon_a$) defined within the Courant–Snyder formalism, for which $\langle g_a\rangle=2\epsilon_{a,\mathrm{rms}}$.

The central task in evaluating Eq.~(\ref{time}) is therefore to calculate the time derivative of the expectation value of the corresponding invariant. Using the SLIM formalism\cite{chao2009slim}, this quantity can be expressed as
\begin{equation}
    \frac{d\langle g_a\rangle}{dt}=\frac{1}{\gamma}\text{Tr}\left(G_{Pa}'\frac{d\Sigma_P'}{dt'}\right)\label{dgdt}
\end{equation}
where $G'_{P a}$ and $\Sigma'_P$ denote the momentum submatrices of $G'_a$ and $\Sigma'$, respectively. Here, $G_a$ and $\Sigma$ are defined in the laboratory frame, whereas $G'_a$ and $\Sigma'$ are defined in the beam rest frame. Their relationship involves a combination of coordinate transformations and a Lorentz transformation, as described below.

The matrix $G_a$ represents one of the three matrices of the quadratic forms of the invariants in the storage ring. Once the lattice configuration is specified, $G_a$ is uniquely determined. For a linear system, these invariants are quadratic forms of the phase space coordinates $\mathbf{z}=(x,x',y,y',z,\delta)$, and are commonly referred to as emittances. They can be written as
\begin{equation}
    g_a=\mathbf{z^T}G_a\mathbf{z},\quad a=x,y,z
\end{equation}
In the decoupled case, the matrix $G_a$ is determined by the corresponding Courant-Snyder parameters. For example,
\begin{equation}
    G_x=\left(
    \begin{matrix}
    \gamma_x&\alpha_x&0&0&0&0\\
    \alpha_x&\beta_x&0&0&0&0\\
    0&0&0&0&0&0\\
    0&0&0&0&0&0\\
    0&0&0&0&0&0\\
    0&0&0&0&0&0\\
    \end{matrix}
    \right)
\end{equation}
Here, $\alpha_x,\beta_x,\gamma_x$ are the horizontal Twiss parameters.

The matrix $\Sigma$ denotes the equilibrium second-moment matrix of the beam distribution in phase space. It is determined by the storage-ring lattice and the corresponding equilibrium beam parameters. Further details can be found in Ref. \cite{chao2020lectures}.

The transformation between the laboratory-frame quantities $(G_a,\Sigma)$ and their rest-frame counterparts $(G'_a,\Sigma')$ consists of two steps:
\begin{enumerate}
    \item A coordinate transformation from $(x,x',y,y',z,\delta)_{\mathrm{lab}}$ to $(x,p_x,y,p_y,z,\delta)_{\mathrm{lab}}$.
    \item A Lorentz transformation from the laboratory frame to the beam rest frame, yielding the coordinates $(x,p_x,y,p_y,z,p_z)_{\mathrm{rest}}$.

\end{enumerate}
For a detailed derivation of these transformations, readers are referred to Ref. \cite{nash2006analytical}.

Previous demonstration shows that in order to evaluate the growth rate with Eq.~(\ref{dgdt}), the key is calculating the term $\frac{d \Sigma'}{dt'}$. Kubo and Oide \cite{kubo2001intrabeam} presented a comprehensive calculation of $\frac{d \Sigma'}{dt'} $for a single-Gaussian beam distribution. Several key results from their work, which will serve as the basis for the generalization to the multi-Gaussian model, are summarized below.

Under the single-Gaussian assumption, the beam distribution in the beam rest frame is given by
\begin{align}
\rho(\mathbf{z})
&=\frac{N}{\Gamma}
\exp\left(-\frac{1}{2}\mathbf{z}^T\Sigma'^{-1}\mathbf{z}\right)
\label{gaussian-exp}\\
&=\frac{N}{\Gamma}
\exp\left[
-\frac{1}{2}\sum_{i,j=x,y,z}
\left(
A_{ij}p_i p_j
+2B_{ij}p_i x_j
+C_{ij}x_i x_j
\right)
\right],\notag
\end{align}
where $\mathbf{z}=(x,p_x,y,p_y,z,p_z)$, and $A, B$, and $C$ are the $3\times3$ submatrices of the inverse covariance matrix, defined by
\begin{equation}
\Sigma'^{-1}=
\begin{pmatrix}
C & B^T\\
B & A
\end{pmatrix}.
\end{equation}
$\Gamma$ is the normalization factor of the distribution. $N$ denotes for the total number of particles in one Gaussian bunch. The normalization factor is defined as 
\begin{equation*}
    \Gamma=\int\exp\left(-\frac{1}{2}\mathbf{z}^T\Sigma'^{-1}\mathbf{z}\right)d^6\mathbf {z}
\end{equation*}
Each binary collision contributes to the evolution of $\frac{d \Sigma'}{dt'}$. The corresponding expectation value can be evaluated by integrating over the phase space distributions of the two colliding particles and the possible final momenta.
\begin{align}
    \left(\frac{d\Sigma_P'}{dt'}\right)_{ij}=&\frac{1}{2N}\iiint\rho_1\rho_2\notag\Delta(p_ip_j)d^3\mathbf{x}d^3\mathbf{p}_1d^3\mathbf{p}_2\notag\\&\int d^3\mathbf{p}_1'\int d^3\mathbf{p}_2'|M_c|^2\frac{m_e^4}{E_1E_2E_1'E_2'}\notag\\&\frac{\delta^4(p'^\mu_1+p_2'^\mu-p_1^\mu-p_2^\mu)}{(2\pi)^2}\label{B-M}
\end{align}
Here, $\rho_1$ and $\rho_2$ are the phase space densities of the two colliding particles, $m$ is the electron rest mass, and $\Delta(p_i p_j)$ denotes the change in the product $p_i p_j$ due to a collision. The quantities $E$ and $E'$ represent the total energies of an electron before and after the collision, respectively, while $p^\mu$ and $p'^\mu$ denote the corresponding four-momenta. The quantity $|M_c|^2$ is the invariant Coulomb scattering amplitude used in the B–M model.

Substituting Eq.~(\ref{gaussian-exp}) into Eq.~(\ref{B-M}), and carrying out some simplification, the result is as follows.
\begin{align}
    \left(\frac{d\Sigma_P'}{dt'}\right)_{ij}&=\frac{2\pi\alpha^2m_eN}{\Gamma}(\text{log})\int\frac{d^3\Delta}{|\Delta|}e^{-\tilde{S}}\left(\delta_{ij}-3\frac{\Delta_i\Delta_j}{|\Delta|^2}\right)\label{dSdt_single}\\
    \tilde{S}&=\Delta^TA\Delta=\sum_{a,b=1}^3A_{ab}\Delta_a\Delta_b\notag
\end{align}
Here, $\alpha$ is the fine-structure constant. $\Delta\triangleq\frac{\mathbf{p_1}-\mathbf{p_2}}{2}$ represents the relative momentum between two collision particles. $(\text{log})$ is the Coulomb log parameter defined by the ratio between the maximum and minimum value of impact parameter $b_{\text{max}},b_{\text{min}}$.
\begin{equation*}
    (\text{log})=\ln\left(\frac{b_{\text{max}}}{b_{\text{min}}}\right)
\end{equation*}
There are various definition of $b_{\text{min}},b_{\text{max}}$, for practical usage, Coulomb log is usually chosen to be a constant between 10 to 20.

The integral in Eq.~(\ref{dSdt_single}) can be further simplified through the techniques developed by Kubo and Oide \cite{kubo2001intrabeam}, ultimately reducing the calculation to three one-dimensional integrals that are straightforward to evaluate numerically. Substituting the resulting momentum second moment evolution rate into Eq.~(\ref{dgdt}) then yields the IBS growth rates of the beam emittances.

\section{Multi-Gaussian model}
A single-Gaussian distribution is often insufficient to describe the phase-space structure of realistic beams. A natural extension is to represent the beam distribution as a superposition of multiple Gaussian components. The double-Gaussian model provides a simple yet physically meaningful starting point for this approach. For example, two Gaussian components with identical covariance matrices but different longitudinal centroids can be superimposed to construct a distribution with a flatter longitudinal profile, thereby providing an approximate representation of flat-top beam distributions that cannot be adequately described by a single Gaussian. Beyond this case, the multi-Gaussian representation is particularly useful for a broader class of non-Gaussian distributions. A representative example is SSMB, where neighboring microbunches may overlap in the longitudinal direction along the storage ring. Motivated by these considerations, we first introduce the double-Gaussian model and then generalize it to the multi-Gaussian case.

Building upon the single-Gaussian model, we extend the beam distribution to a superposition of multiple Gaussian components. To illustrate the underlying principle and facilitate the subsequent generalization, we first introduce the double-Gaussian model.

The phase space distribution of a double-Gaussian beam is defined as the sum of two Gaussian functions sharing the same covariance matrix $\Sigma'$ but having different mean values. Without losing generality, we set the origin at the centroid of two Gaussian components.
\begin{align}
\rho(\mathbf{z}) &=\frac{N}{\Gamma}\left[c_1\exp{\left(-\frac{1}{2}(\mathbf{z} - \mathbf{z}_0/2)^T\Sigma’^{-1}(\mathbf{z} - \mathbf{z}_0/2)\right)}\right.\notag\\
&\quad\left.+c_2\exp{\left(-\frac{1}{2}(\mathbf{z} + \mathbf{z}_0/2)^T\Sigma'^{-1}(\mathbf{z} + \mathbf{z}_0/2)\right)}\right]\label{double-gaussian-exp}
\end{align}
Where $c_1$ and $c_2$ are weighting coefficients satisfying $c_1+c_2=1$, and $\mathbf{z}_0=(x_0,p_{x0},y_0,p_{y0},z_0,p_{z0})$ represents the 6D interval between the centers of two Gaussian distributions. $N$ is the total number of particles in two Gaussian components, $\Gamma$ is the normalization factor of one Gaussian component.

Following a derivation analogous to that presented in Section II, $\left(\frac{d\Sigma_P'}{dt'}\right)_{ij}$ for the double-Gaussian distribution can be expressed in terms of the single-Gaussian result:
\begin{align}
    &\left(\frac{d\Sigma_P'}{dt'}\right)^{\text{double}}_{ij}=\left(c_1^2+c_2^2+2c_1c_2F_{ij}\right)\left(\frac{d\Sigma_P'}{dt'}\right)^{\text{single}}_{ij}\label{final}\\
    &F_{ij}=\frac{\int\frac{d^3\Delta}{|\Delta|}e^{-\tilde{S'}}(\delta_{ij}-3\frac{\Delta_i\Delta_j}{|\Delta|^2})}{\int\frac{d^3\Delta}{|\Delta|}e^{-\tilde{S}}(\delta_{ij}-3\frac{\Delta_i\Delta_j}{|\Delta|^2})}\notag\\
    &c_1+c_2=1\notag\\
    &\tilde{S}'=V^\text{T}\Sigma'^{-1}V\label{tildeS}\\
    &V=\left(-\frac{x_0}{2},\Delta_1-\frac{p_{x0}}{2},-\frac{y_0}{2},\Delta_2-\frac{p_{y0}}{2},-\frac{z_0}{2},\Delta_3-\frac{p_{z0}}{2}\right)
\end{align}
Eq.~(\ref{final}) describes how the relative displacement between the two Gaussian components modifies the momentum second moment evolution rate. The factor $F_{ij}$ accounts for the scattering contributions between particles belonging to different Gaussian components. In general, the resulting integral cannot be reduced to the three one-dimensional integrals available in the single-Gaussian model. Nevertheless, it can still be evaluated numerically. To avoid the apparent singularity at $|\Delta|=0$ during numerical integration, we can introduce spherical coordinates for the relative momentum:
\begin{equation}
\begin{cases}
\Delta_x=|\Delta|\sin\theta\cos\phi\\
\Delta_y=|\Delta|\sin\theta\sin\phi\\
\Delta_z=|\Delta|\cos\theta
\end{cases}
\end{equation}

Eq.~(\ref{final}) provides the general expression of double-Gaussian cases, and it could be naturally generalized to multi-Gaussian model. The multi-Gaussian distribution is expressed as the summation of finite Gaussian components.
\begin{align}
\rho(\mathbf{z}) &=\frac{N}{\Gamma}\left[\sum_{i=1}^m c_i\exp{\left(-\frac{1}{2}(\mathbf{z} - \mathbf{z}_{0i})^T\Sigma^{-1}(\mathbf{z} - \mathbf{z}_{0i})\right)}\right]\label{multi-gaussian-exp}\\
\sum_{i=1}^mc_i&=1\\
\mathbf{z}_{0i}&=(x_{0i},p_{x0i},y_{0i},p_{y0i},z_{0i},p_{z0i})
\end{align}
Here, $m$ denotes the total number of Gaussian components, $N$ is the total number of particles in all components, $\Gamma$ is the normalization factor of one individual Gaussian component, and $\mathbf{z}_{0i}$ is the center of $i$-th Gaussian component.

The calculation of $\frac{d \Sigma'}{dt'}$ for the multi-Gaussian distribution involves all possible pairs of Gaussian components. Its mathematical expression can be written as
\begin{align}
    &\left(\frac{d\Sigma_P'}{dt'}\right)^{\text{m-bunches}}_{ij}=\left(\sum_{p=1}^m c_p^2+
    \sum_{n,k=1,n\neq k}^mc_nc_k F^{nk}_{ij}\right)\notag\\
    &\quad\quad\quad\quad\quad\quad\quad\quad\cdot\left(\frac{d\Sigma_P'}{dt'}\right)^{\text{single}}_{ij}\label{multi-gaussian}\\
    &F^{nk}_{ij}=\frac{\int\frac{d^3\Delta}{|\Delta|}e^{-\tilde{S'}_{nk}}(\delta_{ij}-3\frac{\Delta_i\Delta_j}{|\Delta|^2})}{\int\frac{d^3\Delta}{|\Delta|}e^{-\tilde{S}}(\delta_{ij}-3\frac{\Delta_i\Delta_j}{|\Delta|^2})}
\end{align}
The quantity $\tilde{S'}_{nk}$ is obtained by substituting the 6D separation between the centers of the $n$-th and $k$-th Gaussian components into the quadratic form in Eq.~(\ref{tildeS}). The quantity $\left(\frac{d\Sigma_P'}{dt'}\right)_{ij}^{\text{single}}$ represents $\left(\frac{d\Sigma_P'}{dt'}\right)_{ij}$ for the entire particle population when all particles are described by one single Gaussian distribution.

Before concluding this section, it is important to clarify the motivation and scope of the multi-Gaussian approach. The model is intended as a mathematical approximation of non-Gaussian phase space distributions. It does not necessarily imply that the physical beam consists of multiple overlapping Gaussian bunches. We could use the example of the flat-top distribution here again. Such a distribution could be approximated by a superposition of two Gaussian components and evaluated using the double-Gaussian model, even though no physically distinct Gaussian bunches are present.

In the SSMB scenario discussed in the following section, however, the overlap has a direct physical interpretation: it originates from the longitudinal lengthening of neighboring microbunches. This provides a particularly illustrative example of the multi-Gaussian model and its application. Nevertheless, the underlying phase space distribution is not required to have a multi-Gaussian form.
\section{SSMB scenario}
In this section, we apply the proposed multi-Gaussian model to a representative SSMB storage ring and compare its IBS predictions with those obtained from the modified B–M model. This application provides a quantitative demonstration of how longitudinal overlap between neighboring microbunches can modify the IBS effect.

In the SSMB storage ring considered here, longitudinal microbunching is provided by an optical laser, resulting in a bucket length on the order of the laser wavelength, which is substantially shorter than that of a conventional RF bucket. Combined with an extremely low global phase slippage factor, this configuration enables the formation and storage of ultrashort electron bunches. Consequently, the centers of adjacent microbunches are separated by only one laser wavelength. In the presence of transverse-longitudinal coupling, however, the equilibrium bunch length can vary substantially along the storage ring. In particular, the local bunch length in dispersive regions can become larger than the longitudinal separation between neighboring bunch centers, resulting in significant overlap between adjacent microbunches. Such overlap introduces additional inter-bunch scattering contributions and is expected to enhance the overall IBS effect. 

\subsection{Ring lattice setup}
The storage-ring lattice considered in this study is similar to that presented in Ref~.\cite{Pans_work}, which was designed for SSMB studies. The main lattice and beam parameters adopted in our calculations are summarized in Table~(\ref{parameter1}).
\begin{table}[width=.9\linewidth,cols=3,pos=h]
\caption{Parameters of an SSMB ring.}\label{parameter1}
\begin{tabular*}{\tblwidth}{@{} CCC@{} }
\toprule
Parameters & Units & Value \\
\midrule
Nominal energy & MeV & 400 \\
Circumference & m & 165.62 \\
Natural emittance & nm rad & 1.09 \\
Horizontal/vertical tune & \(\cdots\) & 15.213/13.188 \\
Phase slippage factor & \(\cdots\) & \(5.0 \times 10^{-7}\) \\
Natural energy spread & \(\cdots\) & \(1.35 \times 10^{-4}\) \\
Horizontal/vertical damping time & ms & 429.3/429.3 \\
Longitudinal damping time & ms & 214.7 \\
Coupling & \(\cdots\) & 0.0092 \\
Bunch length at straight & nm & 106.9 \\
Electron number per bunch & \(\cdots\) & 20,000 \\
Bunching system harmonic & \(\cdots\) & 162,227,080 \\
Bunching system voltage & kV & 100 \\
Bunching system wavelength & $\mu$m & 1.021 \\
\bottomrule
\end{tabular*}
\end{table}
\begin{figure}
    \includegraphics[width=0.5\textwidth]{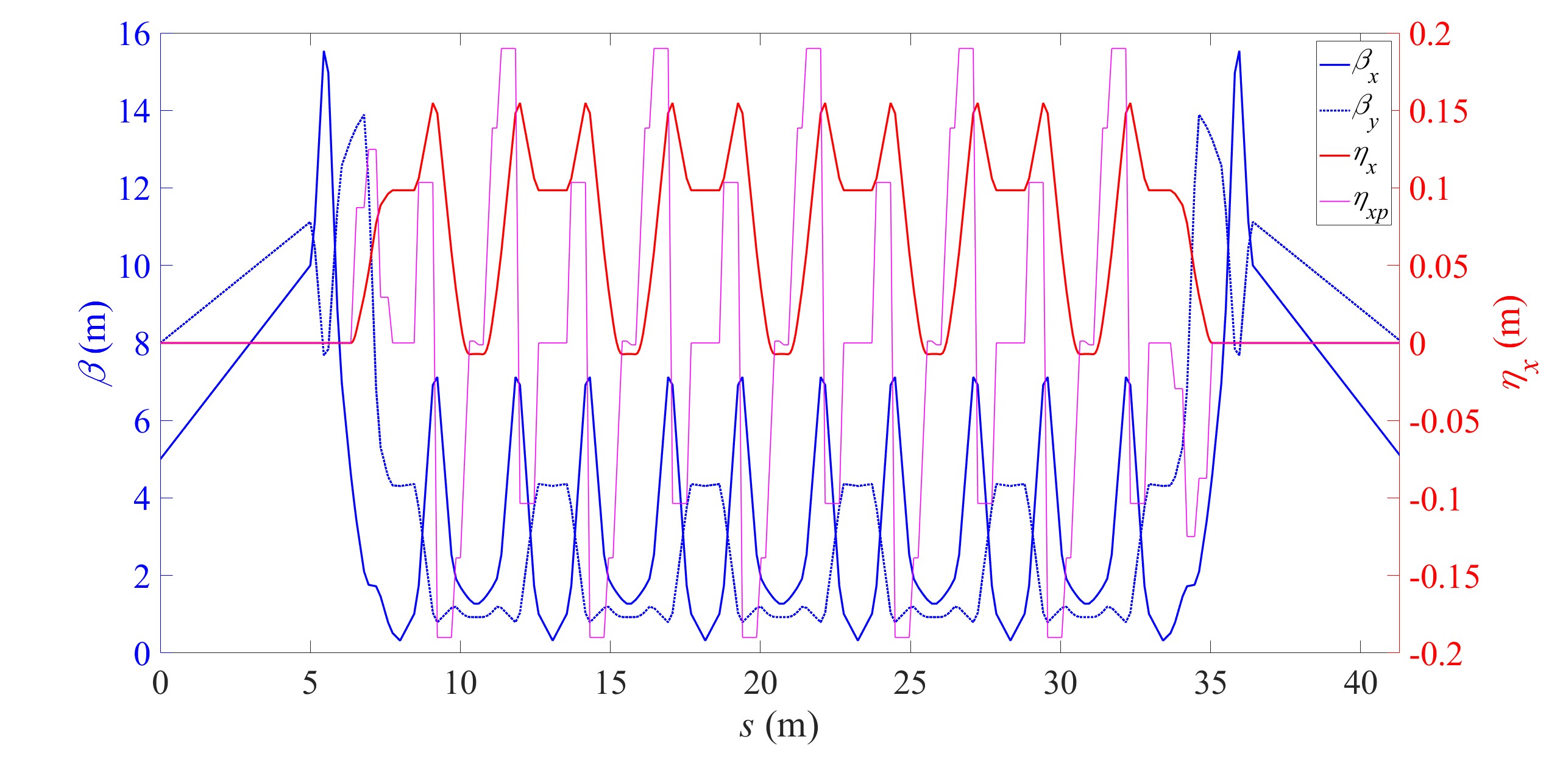}
    \caption{Twiss parameter of the ring}
    \label{twiss}
\end{figure}
\begin{figure}
    \includegraphics[width=0.5\textwidth]{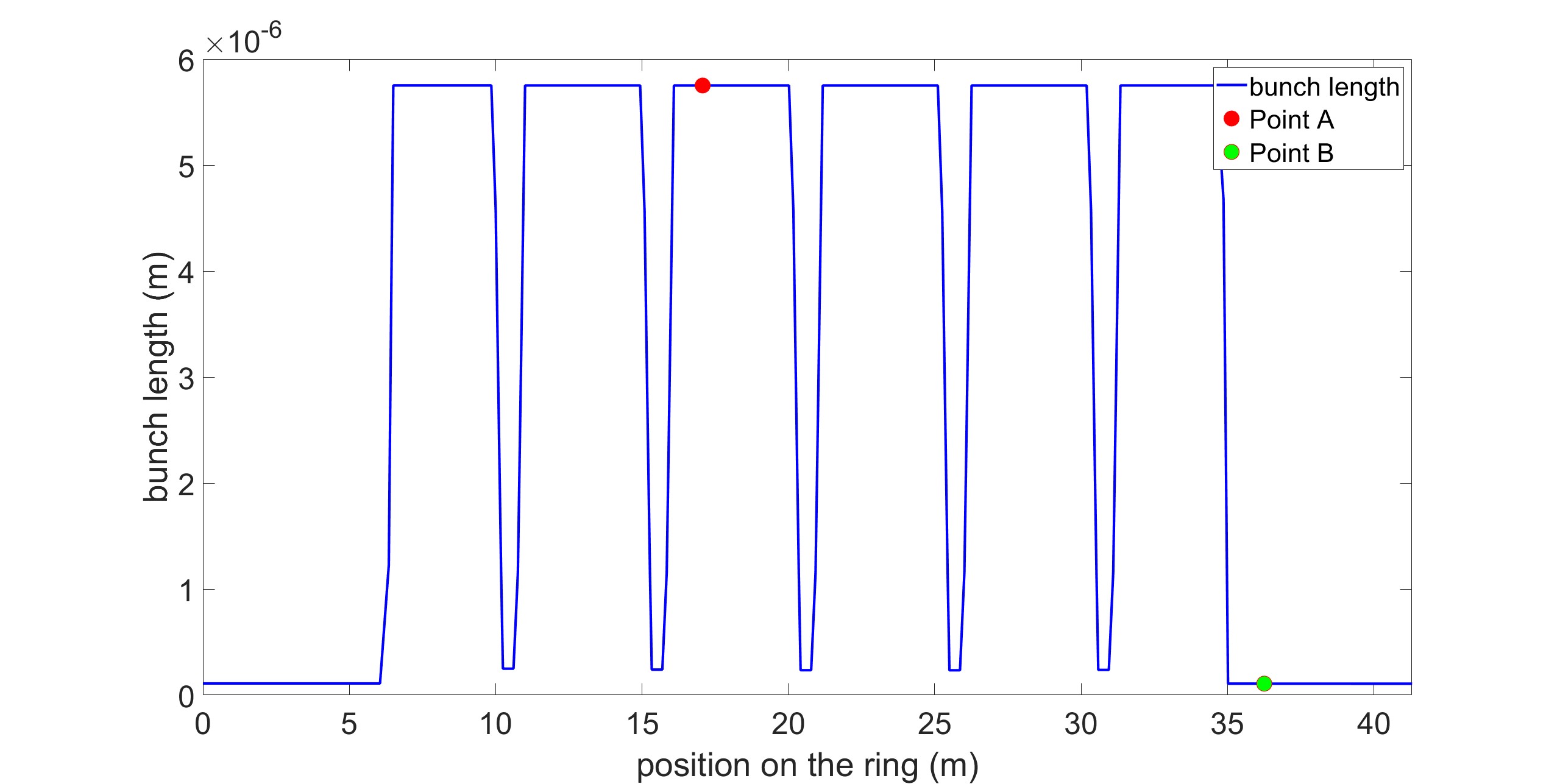}
    \caption{bunch length in one period}
    \label{bunchlength}
\end{figure}

A distinctive feature of this lattice is its ultrashort equilibrium bunch length in the straight sections, which is essential for the stable storage of microbunches. As illustrated in Figure~(\ref{bunchlength}), however, the bunch length varies considerably along the arc sections due to dispersion effects, as also reported in Ref.~\cite{Pans_work}.

The laser wavelength in this lattice is $\lambda=1.021~\mu\mathrm{m}$, which determines the longitudinal separation between adjacent microbunches. Meanwhile, the local equilibrium rms bunch length $\sigma_s$ can exceed $5.7\mu\mathrm{m}$ in certain regions of the ring. Consequently, when $2\sigma_s>\lambda$, the longitudinal distributions of neighboring microbunches can overlap appreciably. In these regions, scattering events are no longer restricted to particles within the same Gaussian component. Collisions between particles belonging to different neighboring components must also be considered. This provides the physical basis for applying the multi-Gaussian model to the SSMB scenario.

\subsection{Gaussian bunch train}
In SSMB scenario, the multi-Gaussian model can be further simplified. We refer to this particular configuration as a "Gaussian bunch train", which consists of multiple Gaussian components equally spaced along the longitudinal direction. All components are assumed to undergo the same dynamical evolution. In this case, the weighting coefficients are equal,
\begin{align}
    &\left(\frac{d\Sigma_P'}{dt'}\right)^{\text{m-bunches}}_{ij}=\left(\sum_{p=1}^m c_p^2+\sum_{n,k=1,n\neq k}^mc_nc_kF^{nk}_{ij}\right)\notag\\&\quad\quad\quad\quad\quad\quad\quad\quad\cdot\left(\frac{d\Sigma_P'}{dt'}\right)^{\text{single}}_{ij}\label{multi-gaussian-nk}\\
    &F^{nk}_{ij}=\frac{\int\frac{d^3\Delta}{|\Delta|}e^{-\tilde{S'}_{nk}}(\delta_{ij}-3\frac{\Delta_i\Delta_j}{|\Delta|^2})}{\int\frac{d^3\Delta}{|\Delta|}e^{-\tilde{S}}(\delta_{ij}-3\frac{\Delta_i\Delta_j}{|\Delta|^2})}\\
    &c_p=\frac{1}{m},p=1,2,\dots,m\notag\\
    &\tilde{S}'_{nk}=V_{nk}^T\Sigma'^{-1}V_{nk},V_{nk}=\left(0,\Delta_1,0,\Delta_2,-\frac{|n-k|\lambda}{2},\Delta_3\right)\label{tildeS}
\end{align}
The relative separation between the centers of the $n$-th and $k$-th components is determined solely by their longitudinal separation, which is an integer times the laser wavelength $\lambda$.

A useful consistency check of this model is obtained by considering the limit in which the separation between neighboring Gaussian components becomes sufficiently large. In this limit, the cross-terms $F^{nk}_{ij}$ vanish, and Eq.~(\ref{multi-gaussian-nk}) reduces to
\begin{equation}
    \left(\frac{d\Sigma_P'}{dt'}\right)^{\text{m-bunches}}_{ij}=\frac{1}{m}\left(\frac{d\Sigma_P'}{dt'}\right)^{\text{single}}_{ij}=\left(\frac{d\Sigma_P'}{dt'}\right)^{\text{1-bunch}}_{ij}
\end{equation}
Here, $\left(\frac{d\Sigma_P'}{dt'}\right)^{\text{1-bunch}}_{ij}$ means the contribution of one Gaussian component to the whole bunch train. Thus its value is $\frac{1}{m}$ of $\left(\frac{d\Sigma_P'}{dt'}\right)^{\text{single}}_{ij}$. which consists of all the particles in the whole bunch train.

This result indicates that the value $\left(\frac{d\Sigma_P'}{dt'}\right)_{ij}$ of $m$ sufficiently separated Gaussian components is identical to that of a single component containing the same total number of particles. This is physically reasonable because, in the absence of inter-component scattering, the IBS contributions arise exclusively from collisions within individual Gaussian components.

For a more practical treatment of SSMB storage rings, it is useful to consider the limit of an infinitely long Gaussian bunch train. In this case, a truncation rule can be introduced to reduce the computational cost of evaluating Eq.~(\ref{multi-gaussian-nk}). More specifically, we seek a cutoff number $m_{\text{cut}}$ beyond which the contribution of additional Gaussian components becomes negligible, leading to the following result.
\begin{equation}
    \left(\frac{d\Sigma_P'}{dt'}\right)^{\text{$m_{\text{cut}}$-bunches}}_{ij}\approx\lim\limits_{m\rightarrow+\infty}\left(\frac{d\Sigma_P'}{dt'}\right)^{\text{m-bunches}}_{ij}
\end{equation}
For simplicity, the proof of the existence of this limit is given in the Appendix. 

In the following, we denote the limit $\lim\limits_{m\rightarrow+\infty}\left(\frac{d\Sigma_P'}{dt'}\right)^{\text{m-bunches}}_{ij}$ as $\left(\frac{d\Sigma_P'}{dt'}\right)^{\text{infinite}}_{ij}$. The term "Infinite" here refers to an idealized Gaussian bunch train consisting of infinitely many equally spaced Gaussian components. Note that the particle number in each Gaussian component is kept constant as $m$ increases.

To determine an appropriate cutoff, one can evaluate the cross-scattering term in Eq.~(\ref{multi-gaussian-nk}) as a function of the maximum separation between component indices, $|n-k|_{\max}=m$, and examine how rapidly the contributions decay with increasing separation. As a practical estimate, we introduce the cutoff rule
\begin{equation}
    m_{\text{cut}} = \left\lfloor \frac{2\sigma_s}{\lambda} \right\rfloor + 1\label{thumb}
\end{equation}
Eq.~(\ref{thumb}) is referred to as the "truncation rule". Physically, it estimates the number of neighboring Gaussian components that is covered by the longitudinal projection of one certain Gaussian component. Beyond this range, the cross-scattering contributions are generally negligible. Nevertheless, the accuracy of this approximation should be checked for specific beam parameters and operating conditions.

Finally, the IBS growth rates can be calculated by substituting the resulting momentum second moment evolution rate into Eq.~(\ref{dgdt}).
\begin{equation}
    \frac{d\langle g_a\rangle}{dt}=\frac{1}{\gamma}\text{Tr}\left(G_{Pa}'\left(\frac{d\Sigma_P'}{dt'}\right)^{\text{infinite}}\right)
\end{equation}
The specialized multi-Gaussian model therefore provides an analytical framework for evaluating IBS in one certain kind of non-Gaussian distributions, and demonstrates the effect of overlapping bunches.
\subsection{Application on SSMB lattice}
\begin{figure}
    \includegraphics[width=0.5\textwidth]{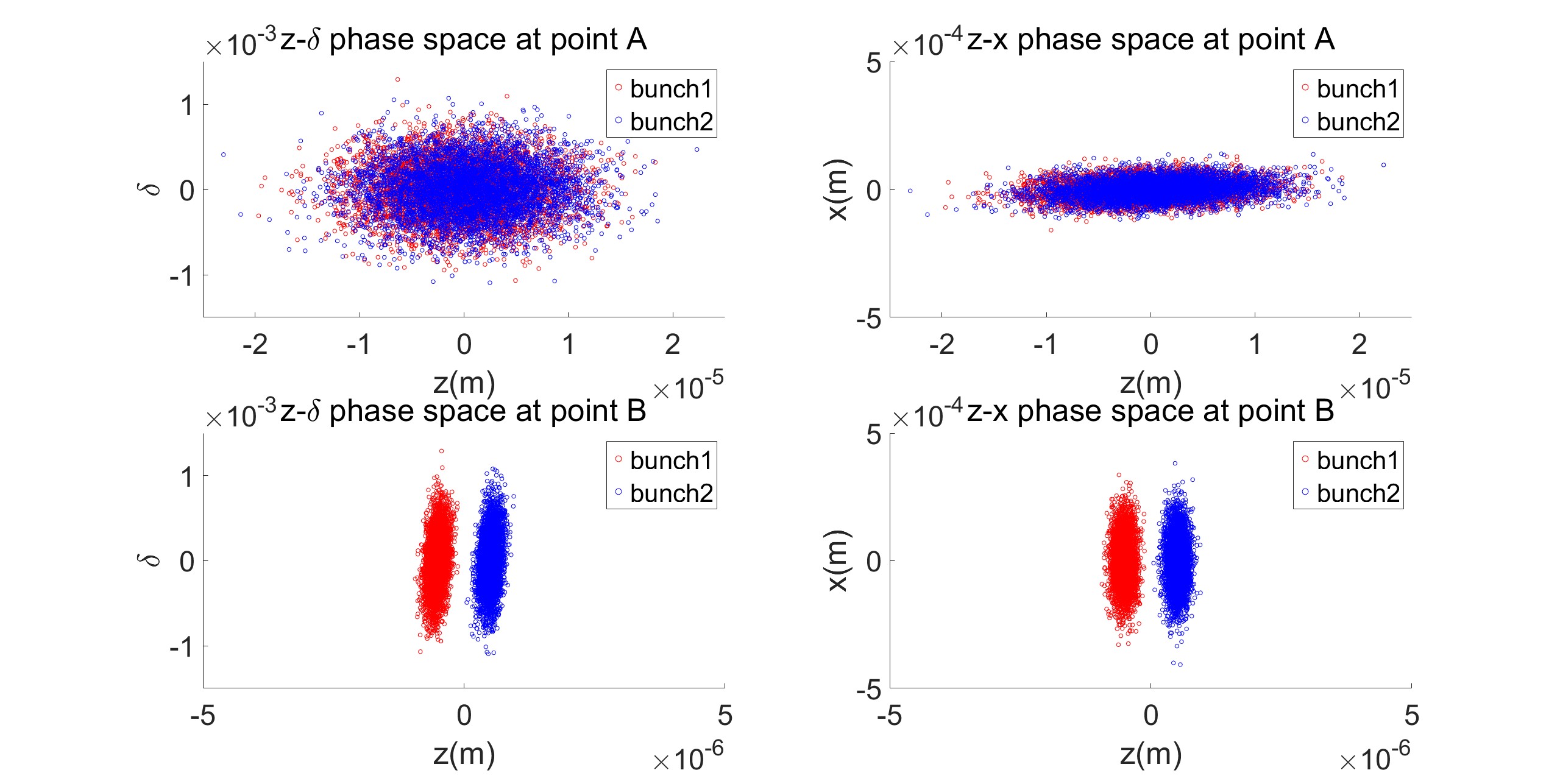}
    \caption{two adjacent bunches distributions at point A\&B}
    \label{phasespace_chap5}
\end{figure}

To quantitatively investigate the IBS enhancement induced by bunch overlap, we examine the beam distributions at two representative locations in the storage ring, denoted as points A and B in Fig.~(\ref{bunchlength}). These locations correspond to regions with substantially different equilibrium bunch lengths. The upper panels of Fig.~(\ref{phasespace_chap5}) show the longitudinal overlap between two adjacent bunches at point A, where the local bunch length is sufficiently large relative to the laser wavelength. In contrast, the lower panels show the well-separated bunch distributions at point B, where the bunch length reaches a minimum.
We then apply the multi-Gaussian model to calculate the IBS growth rates of an infinite Gaussian bunch train. Using the SLIM formalism, we obtain the second moment matrix at each lattice element and determine the number of Gaussian components according to the truncation rule derived previously. In the present calculation, $m=12$ components are retained. 

\begin{figure}
    \centering
    \begin{subfigure}{0.45\textwidth}
        \centering
        \includegraphics[width=\linewidth]{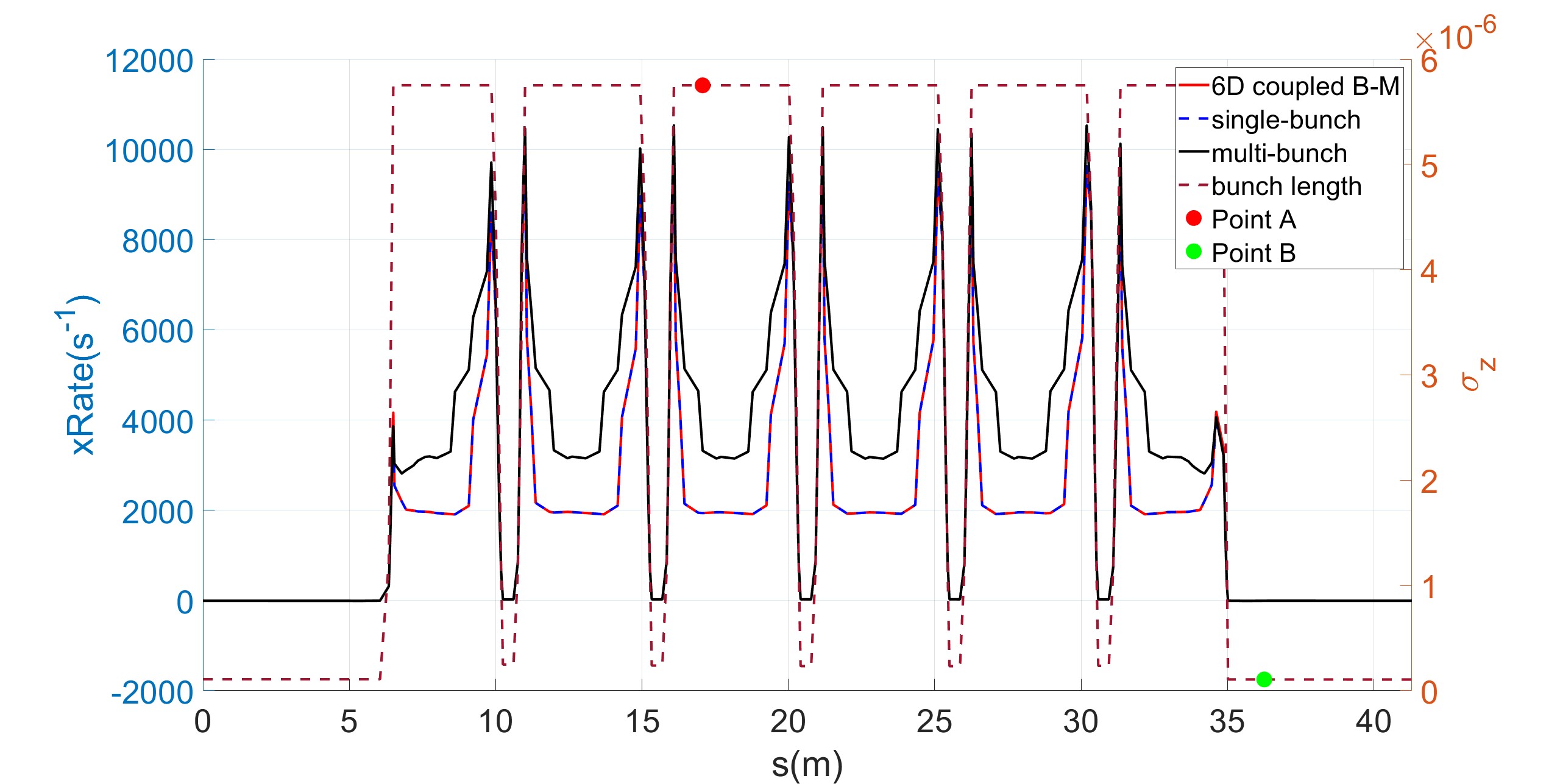}
        \caption{Horizontal IBS growth rate}
        \label{fig:ibs_horizontal}
    \end{subfigure}
    \hfill
    \begin{subfigure}{0.45\textwidth}
        \centering
        \includegraphics[width=\linewidth]{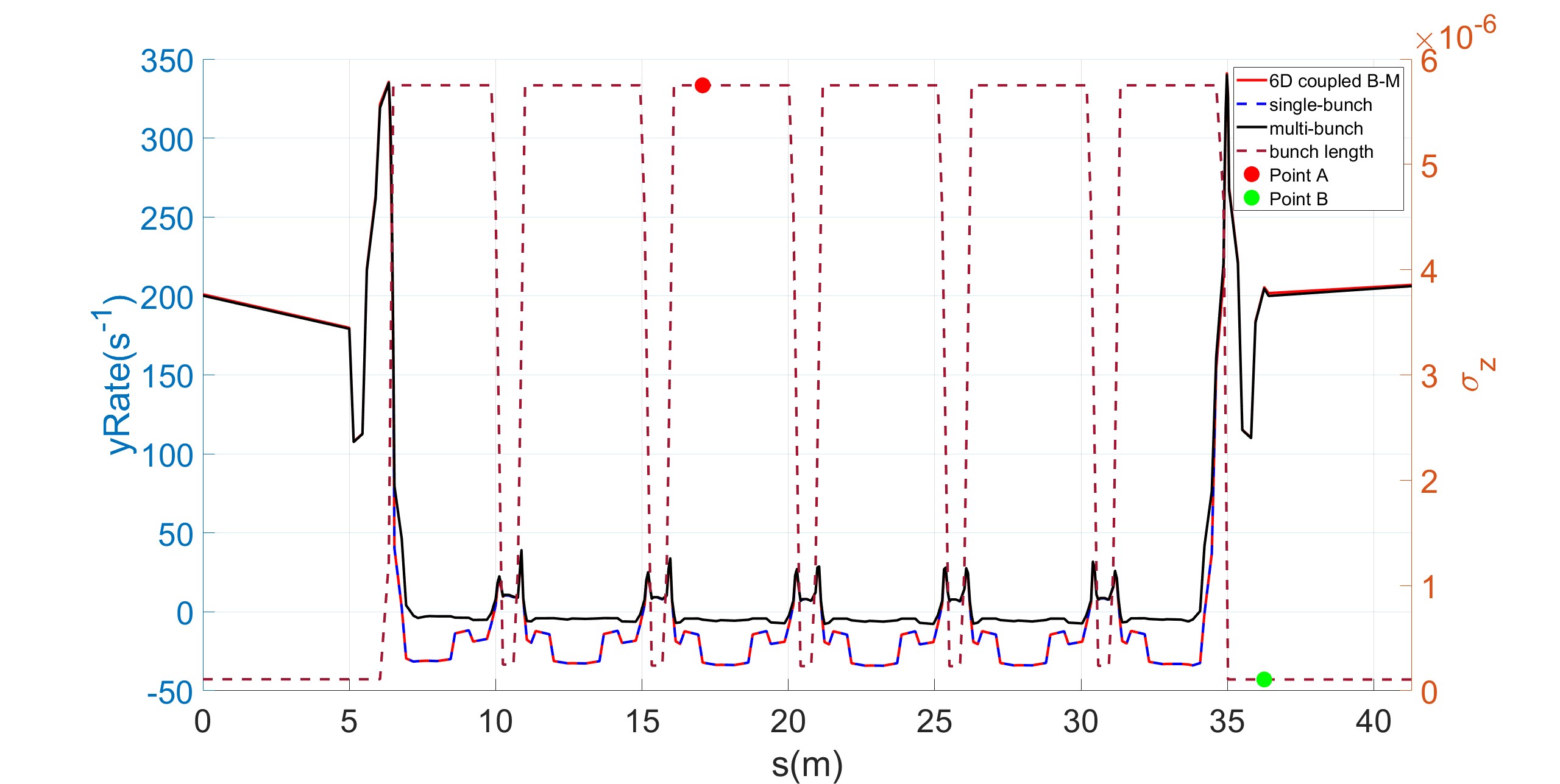}
        \caption{Vertical IBS growth rate}
        \label{fig:ibs_vertical}
    \end{subfigure}
    \hfill
    \begin{subfigure}{0.45\textwidth}
        \centering
        \includegraphics[width=\linewidth]{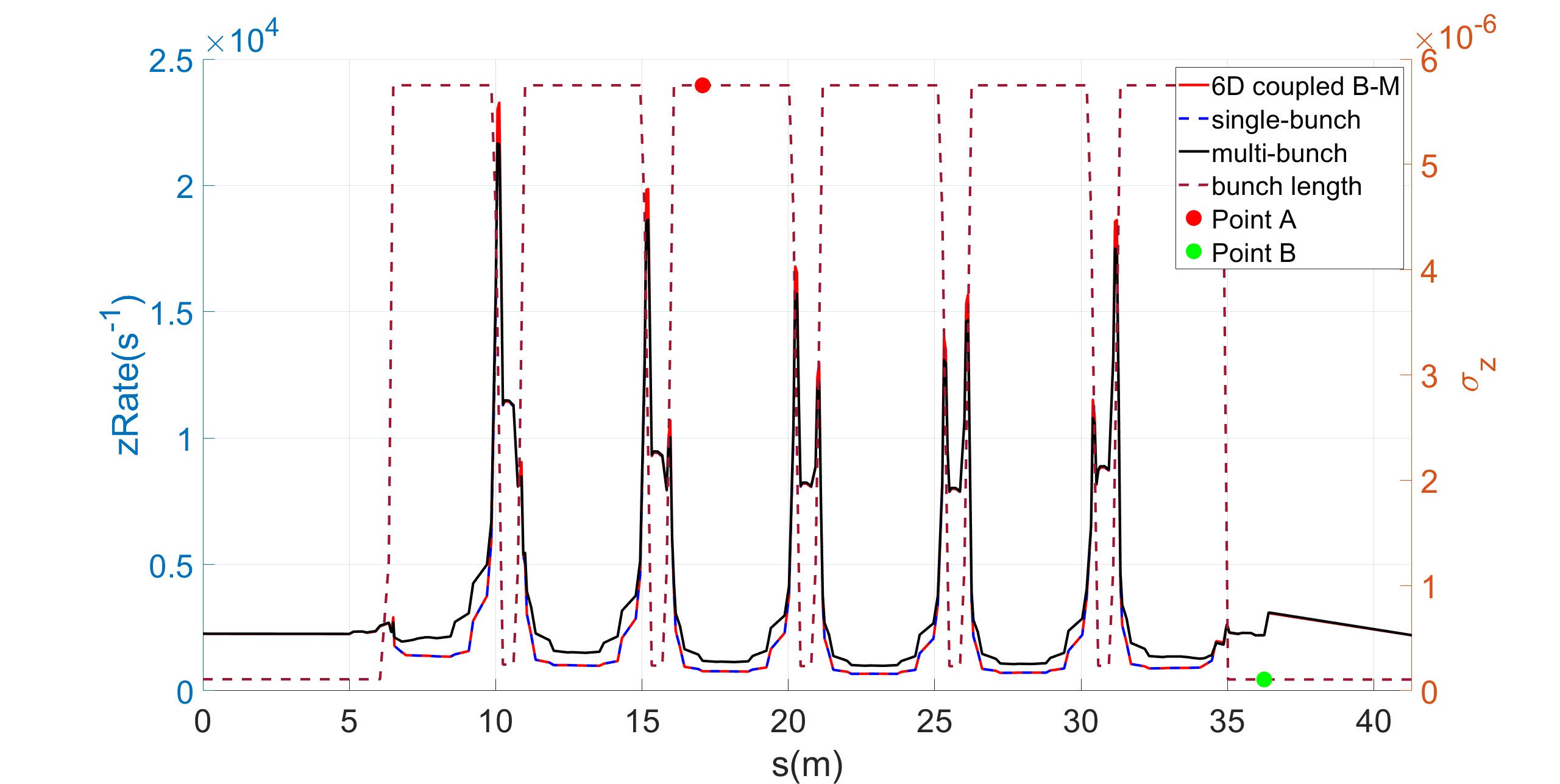}
        \caption{Longitudinal IBS growth rate}
        \label{fig:ibs_longitudinal}
    \end{subfigure}
    
    \caption{IBS growth rates comparison between different model}
    \label{fig:ibs_growth_rates}
\end{figure}

We applied the multi-Gaussian model to calculate the "first-turn IBS" of the whole SSMB storage ring, and Figure.~\ref{fig:ibs_growth_rates} illustrates our results. 

The "first turn IBS" is a concept opposite to the "equilibrium IBS". "first turn IBS" means evaluating the IBS growth rate under given emittances, while "equilibrium IBS" means iteratively solving the equilibrium equation, consisting of quantum excitation, radiation damping, and IBS. Here we adopted the emittances given in Tab.~\ref{parameter1}, thus the results are "first turn IBS".

As expected, the single-Gaussian model reproduces exactly the results of the modified B–M model. This agreement provides a consistency check for the multi-Gaussian formulation, confirming that it recovers the conventional IBS model in the single-Gaussian limit. In contrast, the multi-Gaussian results deviate from those of the conventional model in regions where the local bunch length becomes comparable to or larger than the longitudinal separation between adjacent microbunches. The IBS growth rates increase in all three directions, indicating that the additional scattering contributions associated with bunch overlap enhance the overall IBS effect.

The SSMB scenario therefore provides a representative example in which the differences between Gaussian and non-Gaussian IBS models become particularly pronounced. The proposed multi-Gaussian model offers a practical framework for evaluating these differences and provides further insight into how non-Gaussian phase space distributions can modify IBS-induced beam evolution.
\section{Conclusion and extension}
In this paper, we have extended the conventional single-Gaussian IBS model to double- and multi-Gaussian distributions, thereby generalizing analytical IBS calculations to a broader class of non-Gaussian beam distributions. 

The multi-Gaussian model allows certain kinds of non-Gaussian distributions to be approximated through a superposition of Gaussian components, providing a means of evaluating the corresponding IBS effect. Furthermore, we have established a limiting formulation for an infinite Gaussian bunch train, together with a practical truncation rule, to quantify the enhancement of IBS caused by overlapping microbunches.

To demonstrate the applicability of the proposed model, we applied the multi-Gaussian model to an SSMB storage ring. The results show that longitudinal overlap between neighboring microbunches can substantially modify the IBS growth rates. Based on the proposed model, we have established a computational approach for evaluating IBS in SSMB storage rings while accounting for the additional scattering contributions associated with bunch overlap. These findings provide a reference for assessing IBS effect in SSMB and other alternative storage ring designs. 

Compared with the longitudinal-slicing method implemented in codes such as ELEGANT~\cite{xiao2015intra}, the multi-Gaussian approach offers a different perspective on representing non-Gaussian beam distributions. The spirit of slicing technique is breaking the non-Gaussian distribution into infinitesimal Gaussian slices. By contrast, the multi-Gaussian model considers the non-Guassian distribution as a whole. It could reveal some of the fundamental differences between Gaussian distributions and non-Gaussian distributions.

Despite these advances, the multi-Gaussian model does not fully describe the evolution of the beam distribution under IBS. In fact, it only describes the second moment evolution. This is also the limitation of conventional IBS analytical model such as Piwinski and B-M.

A more general framework for describing the IBS-induced evolution of arbitrary phase space distributions is provided by the Fokker-Planck equation, following the approach developed by Zenkevich et al. \cite{zenkevich2006new} Within this framework, the conventional description in terms of emittances growth can be extended to the evolution of the full phase space distribution and its associated eigenmodes. The IBS effect is characterized by the drift coefficients $D_i$ and diffusion coefficients $D_{ij}$, corresponding to the first and second-order Kramers-Moyal coefficients, respectively. The Zenkevich approach provides a more comprehensive description of IBS and offers a natural framework for incorporating its interplay with other dynamical effects in accelerator physics. Following a derivation similar to that presented in Section III, the multi-Gaussian model can be used to evaluate the phase space averaged diffusion coefficients, $\langle D_{ij}\rangle$, which can then be incorporated into the Zenkevich approach to account for IBS effect associated with non-Gaussian beam distributions.

The multi-Gaussian model presented in this work can therefore be regarded as an intermediate step toward a more general theory of non-Gaussian IBS. Further development along this direction may establish a framework for describing the self-consistent evolution of non-Gaussian beam distributions under IBS.

\section{Acknowledgments}
The authors would like to thank Alexander Wu Chao, Jingyuan Zhao, and Jiazhen Tang for helpful discussions. This work is supported by the National Key R\&D Program of China (Grant No.2022YFA1603400), the Beijing Natural Science Foundation (Grant No.1254058), the National Natural Science Foundation of China (NSFC Grant No.12035010), and the Tsinghua University Initiative Scientific Research Program.

\appendix
\section{Proof of convergence of $\left(\frac{d\Sigma_P'}{dt'}\right)^{\text{infinite}}_{ij}$}
\label{app:convergence}

This appendix proves the convergence of the summation in $\left(\frac{d\Sigma_P'}{dt'}\right)^{\text{m-bunches}}_{ij}$ in the limit of an infinite bunch train ($m \to +\infty$). The $l$-th term in the sum can be expressed as
\begin{align}
    a_l &= 2\frac{m-l}{m} \cdot \frac{G(l\lambda)}{G(0)}, \label{eq:al_def} \\
    G(l\lambda) &:= \int \frac{d^3\Delta}{|\Delta|} \,
        e^{-\tilde{S}'_{nk}} \,
        \Bigl(\delta_{ij} - 3\frac{\Delta_i\Delta_j}{|\Delta|^2}\Bigr), \label{eq:F_def} \\
    \tilde{S}'_{nk} &:= V_{nk}^T \Sigma^{-1} V_{nk}, \qquad
        V_{nk} = \bigl(0,\,\Delta_1,\,0,\,\Delta_2,\,-\frac{|n-k|\lambda}{2},\,\Delta_3\bigr). \nonumber
\end{align}
Note that $|n-k| = l$ in this context.

We prove absolute convergence. First, consider the absolute value of the general term:
\begin{align}
    |a_l|
    &= 2\left| \frac{m-l}{m} \right| \,
       \left| \frac{G(l\lambda)}{G(0)} \right| \notag \\
    &< 2\left| \frac{G(l\lambda)}{G(0)} \right| \qquad (\text{since } |(m-l)/m| < 1 \text{ for } l \geq 1). \label{eq:ineq1}
\end{align}

The function $G(l\lambda)$ can be bounded by exploiting the positive-definiteness of $\Sigma^{-1}$. Let $t_{\min} > 0$ be the smallest eigenvalue of $\Sigma^{-1}$. Then,
\begin{equation}
    \tilde{S}'_{nk} = V_{nk}^T \Sigma^{-1} V_{nk}
    \;\geq\; t_{\min} \, V_{nk}^T V_{nk}
    \;=\; t_{\min} \bigl( |\Delta|^2 + \frac{(l\lambda)^2}{4} \bigr). \label{eq:bound_S}
\end{equation}
Using this inequality, we obtain an upper bound for $|F(l\lambda)|$:
\begin{align}
    \bigl| G(l\lambda) \bigr|
    &\leq\int\frac{d^3\Delta}{|\Delta|}e^{-\tilde{S'}_{nk}}\left|(\delta_{ij}-3\frac{\Delta_i\Delta_j}{|\Delta|^2})\right|\\
    &\leq \int \frac{d^3\Delta}{|\Delta|} \,
        e^{-t_{\min}[|\Delta|^2 + \frac{(l\lambda)^2}{4}]} \,
        \Bigl| \delta_{ij} - 3\frac{\Delta_i\Delta_j}{|\Delta|^2} \Bigr| \notag \\
    &= e^{-t_{\min} \frac{(l\lambda)^2}{4}} \,
        \underbrace{\int \frac{d^3\Delta}{|\Delta|} \,
        e^{-t_{\min}|\Delta|^2} \,
        \Bigl| \delta_{ij} - 3\frac{\Delta_i\Delta_j}{|\Delta|^2} \Bigr|}_{=: \, C_0}. \label{eq:bound_F}
\end{align}
The integral $C_0$ is a finite, positive constant independent of $l$.

Substituting the bound \eqref{eq:bound_F} into \eqref{eq:ineq1} yields
\begin{equation}
    |a_l| \;<\; \frac{2C_0}{|G(0)|} \; e^{-t_{\min} \frac{(l\lambda)^2}{4}}
    \;=:\; C \, e^{-t_{\min} \frac{(l\lambda)^2}{4}}, \label{eq:final_bound}
\end{equation}
where $C =: 2C_0 / |G(0)|$ is a constant.

It is well-known that the series $\sum_{l=1}^{\infty} e^{-t_{\min} \frac{(l\lambda)^2}{4}}$ is a convergent series. By the Direct Comparison Test, the inequality \eqref{eq:final_bound} implies that $\sum_{l=1}^{\infty} |a_l|$ converges. Absolute convergence of $\sum_{l=1}^{\infty} a_l$ follows, which guarantees the convergence of the original summation in $\left(\frac{d\Sigma_P'}{dt'}\right)^{\text{m-bunches}}_{ij}$ as $m \to +\infty$.

\bibliographystyle{elsarticle-num}

\bibliography{cas-refs}

\end{document}